# A Dynamical Model for Information Retrieval and Emergence of Scale-Free Clusters in a Long Term Memory Network

Ignazio Licata

[1] ISEM, Institute for Scientific Ethics and Methodology, Palermo, Italy
Ignazio.licata@ejtp.info

*Abstract*

*The classical forms of knowledge representation fail when a strong dynamical interconnection between system and environment comes into play. We propose here a model of information retrieval derived from the Kintsch-Ericsson scheme, based upon a long term memory (LTM) associative net whose structure changes in time according to the textual content of the analyzed documents. Both the theoretical analysis carried out by using simple statistical tools and the tests show the appearing of typical power-laws and the net configuration as a scale-free graph. The information retrieval from LTM shows that the entire system can be considered to be an information amplifier which leads to the emergence of new cognitive structures. It has to be underlined that the expanding of the semantic domain regards the user-network as a whole system.*

*Key- words:* Systemic Approach; Intelligent Agent; Knowledge Dynamical Representation; Long Term Memory; Scale-Free Graphs

## 1. Introduction

The concept of an intelligent agent involves a world description by which the agent makes its choices, activating some data evaluation strategies and selecting the most significant elements on the basis of a given objective and the interaction with the environment.

The old strong AI approaches were based on formal logic tools and heuristic rules for achieving a sufficiently exhaustive world description.

Expert Systems used in various fields – chess, military and economic strategies, clinical diagnosis etc. – even though they used different formal tools for conceptualization, they were all classifiable within the strong AI and they shared a static knowledge representation form.

A world description made up by a set of well defined and ever accessible production rules inevitably leads to a limited efficacy as the semantic domain is expanding

Connectionistic approaches based on neural networks and parallel distributed processing systems have opened new prospects characterized by a different and closer relation between the system and the environment.

The new systemic-cybernetic approach by N. Wiener, L. Von Bertalanffy, R. Ashby, H. Von Foerster,, H. Maturana and F. Varela makes a world description depending on the observer. This is considered to be an interacting system that selects knowledge according to its inner structure and its dynamic history.

Between the observer (the intelligent system) and the observed (the world) there is no longer a linear and deductive relationship defined by a formal model, but there is a circular function based on a continuous adaptation and coevolution strategy.

This implies at least two important differences from the approaches adopted in strong AI.

A relationship of thermodynamic openness between the system and the environment must be considered. Information and energy make up a flow that crosses and continuously modifies their relationships and their structure.

This dissipative feature of the system is only a necessary but not sufficient condition. The system must also have a logic openness i.e. it must show emergent behaviours which depend on the dynamic state of the relationship between the system and the environment. These emergent behaviours should lead the system to more complex adaptive situations by the production of new knowledge.

So there are no longer definite and independent knowledge representation forms of the agent-observer world description. But the agent continuously generates adaptive processes that produce at any moment a world description bound to the structure and the objectives of the agent.

Classic AI representations conform to the adoption of models with a low logic openness. They can be seen as samples of a knowledge structure that is always dynamic in the natural systems.

Being aware of these limits and problems we are searching for common aspects between the symbolic and the connectionistic approaches.

Firstly we consider an ontology as a linguistic theory of beings. This implies two strong assumptions:

1) the "neutrality" of the ontological representation. This is formalized by the second order logic and it

must guarantee the system portability i.e. the possibility to make different conceptualizations.
2) The possibility to describe the world by the discrete units of the second order logic at any moment.

Furthermore a knowledge acquisition system must exhibit coherence, operational closure and must produce a world by intrinsic emergent processes based on the dialogic relationship between the system and the user.
So we must leave the ontological models for the ontogenetic models of knowledge representation.

## 2. A Dynamic Knowledge acquisition System

During the discourse comprehension it has been demonstrated that the human mind generates dynamic structures that are adapted to the particular context of use.
In particular Kintsch (1998) noticed that human knowledge in these cases can be modelled by networks of propositions.
This particular model extends and combines the advantages of the classic static forms of representation such as "associative networks" (Meyer and Schvaneveldt, 1971), "semantic networks" (Collins and Quillian, 1969), "frames" (Minsky, 1975) and "scripts" (Schank and Abelson, 1977). They are all based on the predicate-argument schema i.e. their atomic parts are propositions linked to each other by weighted and not labelled arcs.
This particular representation form presents great advantages in comparison to classic formalisms. While semantic networks, frames and scripts organize knowledge in a more ordered and logical way, the networks of propositions are disorganized and chaotic, but their structure can change dynamically in time on the basis of cumulated experiences and the knowledge can be retrieved with particular procedures that consider the perceived context.
The meaning of a node, i.e. its correspondent concept, is given by its position in the network because it is constrained by the close nodes. But during the comprehension process only a part of the neighborhood is considered. This subset of nodes is "activated" by the particular context of use represented by other nodes.
So the meaning of every concept in the network is not static, i.e. fixed and permanent, but it is always built in a particular structure called "working memory" by the activation of a particular section of the network.
To specify the activation modalities Kintsch and Ericsson (Kintsch, Patel, and Ericsson, 1999) introduced the concept of "Long Term Working Memory" (LTWM). This is a part of the "Long Term Memory" (LTM) that is the entire network of propositions. LTWM is generated by the short term part of the working memory (STWM). This process is allowed by fixed and stable memory structures called "retrieval cues" that link the objects present in the STWM to other objects present in the LTM.
After the definition of their model Kintsch and Ericsson have tried to implement it. Two particular problems had to be solved, the creation of the LTWM and the formation of the retrieval cues.
For the definition of the LTWM Kintsch developed two methods. The first, defined with Van Dijk (van Dijk and Kintsch, 1983), is a manual technique that starts from the propositions present in the text (micropropositions) and using some organizing rules arrives to the definition of "macropropositions" and "macrostructures" and even to the definition of LTWM.
The second is based on the "Latent Semantic Analysis" (LSA) (Landauer, Foltz, and Laham, 1998). This technique can infer, from the matrix of co-occurrence rates of the words, a semantic space that reflects the semantic relations between words and phrases. This space has typically 300-400 dimensions and allows to represent words, phrases and entire texts in a vectorial form. In this way the semantic relation between two vectors can be estimated by their cosine (a measure that according to Kintsch can be interpreted as a correlation coefficient).
In the second solution the modalities of the information retrieval from the semantic space are not clearly specified. When the "textbase", i.e. the representation obtained directly from the text, is sufficiently expressed, the retrieval of knowledge from the LTM is not necessary. In other cases a correct comprehension of the text (or the relative "situation model") requires the retrieval of knowledge from the LTM.
After the creation of the LTWM the integration process begins, i.e. the activation of the nodes correspondent to the meaning of the phrase. Kintsch uses a diffusion of activation procedure that is a simplified version of the one developed by McClelland and Rumelhart (McClelland and Rumelhart, 1986). Firstly an activation vector is defined whose elements are indexed over the nodes of LTWM. Any element's value is "1" or "0" depending on the presence or the absence of the corresponding node in the analyzed phrase (i.e. in the STWM). This vector is multiplied by the matrix of the correlation rates (the weights of the links of the LTWM) and the resulting vector is normalized. This becomes the new activation vector that must be multiplied again by the matrix of the correlation rates. This procedure goes on until the activation vector becomes stable. After the integration process, the irrelevant nodes are deactivated and only those that represent the situation model remain activated.
There is also another problem that must be considered. Theorically the position occupied by a word in the LTWM is determined by a lifetime experience, i.e. by the continuous use that is made of it. Obviously this kind of knowledge cannot be reached practically and Kintsch build his semantic space using information taken from a dictionary. It is important to note that this operation is done only one time and the semantic space is not further updated. So this kind of implementation loses some of its dynamic properties.

## 3. An Alternative Implementation of the LTWM Model

We think that all the previous problems can be fully solved only by dropping the intermediate representation of the semantic space and trying to find a method that define directly the retrieval cues. Unfortunately the lack of adequate textual parsers able convert the paragraphs of a text on the correspondent atomic propositions has driven us to develop simple dynamic models of associative networks that are based on the LTWM model of Kintsch and Ericsson.

Our associative network can be extended, in a second phase, using the relations provided by an existing general ontology. In this way we think that can be derived a network of propositions.

We are considering a particular ontology built over WordNet (A. Gangemi, N. Guarino, C. Masolo and A.Oltramari, 2002). Each concept is represented by a synset and is linked to other concepts by relations (propositions). A synset is a set of synonims that represent the same concept (Miller. G.A. , 1993).

There are many systems for the ontology extraction (Navigli, Velardi, and Gangemi, 2003) or for the semantic disambiguation that are based on WordNet. In particular we are considering a technique (Ide N. and Veronis J., 1998) that is based on the extension of the nodes representing the terms of a text (phrase) with all the correspondent synsets of WordNet.

Once the correct synsets have been activated with the diffusion of an activation signal, the relative relations can be added to the structure.

In this way can arise a dynamic structure made by propositions and literals that can be represented in an ontological way, maybe using the formalisms of the Semantic Web Languages (RDF, RDF Schema).

This ontology is created dynamically considering the context i.e. the content of the analyzed text and the nodes representing the goals of the user (query).

The coherence of the ontology is guaranteed by the coherence of the upper level ontology that is used during its assemblage.

Now we are facing the problem of the creation of the associative network that is the first step towards the automatic generation of ontologies.

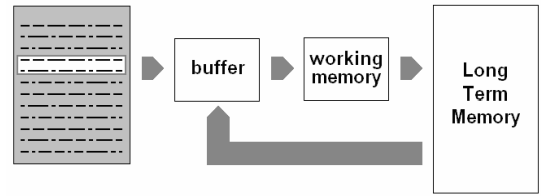

**Figure 1. A possibile architecture of a system for the dynamical acquisition of knowledge from a repository of documents.**

Fig. 1 describes the system that we used for the extraction of knowledge from textual documents. The part of the document that is analysed – a section, a paragraph or a simple group of words enclosed in a window – is stored temporarily in a buffer. Its content must be codified on the basis of the context before being elaborated by the working memory block. This has been implemented by a simple scale-free graph model (Albert and Barabasi, 2001), due to the fact that recently it has been found that human knowledge seems to be structured in this way (Steyvers and Tenenbaum, 2001).

The analysis is performed over all the paragraphs transferring their content in the buffer. This structure contains not only the words of the analyzed paragraph, but also words retrieved from the LTM by the diffusion of an activation signal starting from the LTM nodes that represent words in the buffer. Theorically the buffer should contain also words activated during the analysis of the previous paragraph, but this aspect of Kintsch model has not been considered for its computational complexity.

The buffer, the working memory and the activated part of the LTM block can be compared (but they are not the same structure) to the LTWM defined by Kintsch and Ericsson.

During the acquisition of the paragraph is used a stoplist of words that must not be considered (as articles, pronouns etc.). For any word in the text, the paragraphs where it has appeared (or where it has been inserted after the retrieval procedure) are stored. When the entire text has been parsed and the data of all the N not filtered words have been memorized, the formation of the network of concepts in the working memory begins.

The model adopted is similar to the one defined by Bianconi and Barabasi (2001). The process starts with a net consisting of N disconnected nodes. At every step t=1..N each node (associated with one of the N words) establishes a link with other M units (M=5). If j is the selected unit, the probability that this node establishes a link with the unit i is:

$$P_i = \frac{U_i k_i}{U_1 k_1 + ... + U_N k_N}$$

where $k_i$ is the degree of the unit i[1], i.e. the number of links established by it, while $U_i$ is the fitness value associated to the node that can be computed as the ratio between the number of paragraphs that contain both words i and j and the number of paragraphs that contain either i or j.

LTM is an associative network that is updated with the content of the WM. Whenever a link of the WM corresponds to a link present in the LTM, the weight of this one is increased by "1". For example if the WM links "market" to "economy" and in the LTM "market" is linked to "economy" with weight "7" and to "stock" with weight "4", in the updated LTM "market" is linked to "economy" with weight "8" and to "stock" with weight "4" (unchanged). To perform the diffusion of the activation signal all the weights must be normalized. In this case "market" must be linked to "economy" with weight 8/(8+4) and to "stock" with weight 4/(8+4).

Since the scale free network that represents the content of the WM is used to update the content of LTM, this associative networks should take the form of a scale free graph. Unfortunately the modalities of evolution of the LTM does not allow the definition of a simple equivalent mathematic model, that is necessary to make useful previsions about its evolution.

While in the scale free graph models proposed by literature at each temporal step M are added new nodes to the graph, with M defined beforehand, in the system that we have developed, after the analysis of a new document the links related to an unknown number of nodes of the LTM network are updated on the basis of the content of the WM. This is the number of the words that have not been filtered by the stoplist. Another important difference with other scale free models presented in literature (Dorogovtsev and Mendes, 2001) is the particular fitness function that is used. This function does not depend on a single node but on the considered pair of nodes, i.e the fitness value of a word is not constant but depends on the other word that is linked to it. For example the noun "car" should present for the link with "engine" a fitness value greater than the one presented for the link with "economy".

## 4. Evaluation of the Scale-Free Properties

To test the validity of the scale free graph model adopted for the WM, we gave 100 files of the Reuters Corpus[1] as input to the system disabling the retrieval of information from the LTM. Two versions of the model have been tested, one with bidirectional links and the other with directed links (in this case we considered $k_i = k_{i(IN)} + k_{i(OUT)}$).

In Fig. 2 it is represented an example of a network with bidirectional links. The economic bias of the articles justifies the presence of hubs as "interest rate", "economy", etc., while other frequent words as "child", "restaurant", etc. establish less link with the others.

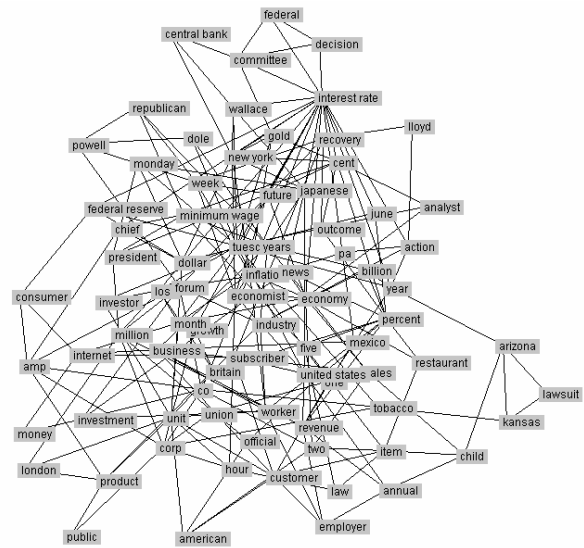

**Figure 2. A network with bidirectional links obtained by the analysis of 100 files of the Reuters Corpus.**

The following graphs report the average path length between each pairs of nodes (Fig.3) and the clustering coefficients (Fig.4). Some networks with different sizes have been considered filtering less or more words in the analysed texts.

All the results seem to confirm those reached by Bianconi and Barabasi. The trend of the average path related to random graphs having the same dimensions of the considered scale free graphs has an higher slope. The clustering coefficient of the scale free graph model has an higher order of magnitude in comparison with the one computed for random networks.

---

[1] Each node is connected to itself by a loop.

[1] Reuters Corpus, Volume 1, English language, 1996-08-20 to 1997-08-19, http://about.reuters.com/researchandstandards/corpus.

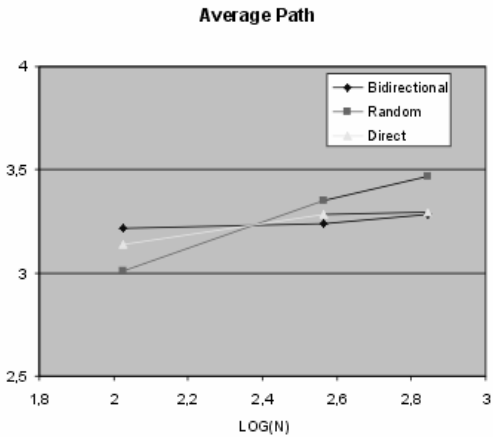

Figure 3. Comparison of average path lengths of different types of networks.

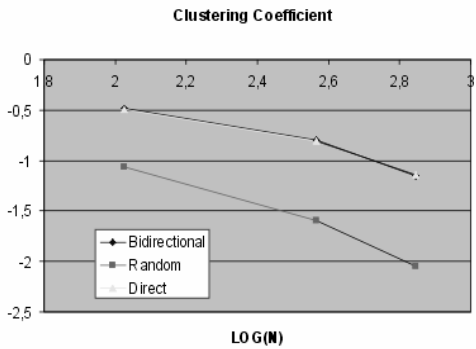

Figure 4. Comparison of clustering coefficients of different types of networks.

Fig. 5 and Fig. 6 show the degree distribution of a graph with bidirectional links and directed links respectively. In the first case the degrees distribution decays as $P(k) \approx k^{-G}$ with G = 3.2657. In the second case the power law trend has a coefficient G = 2.3897.

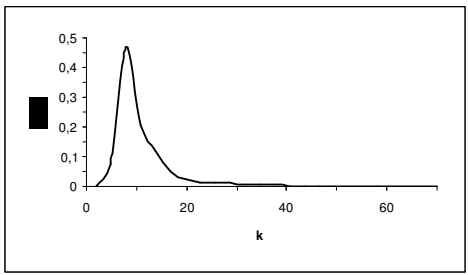

Figure 5. Degree distribution of a graph with M=5 and bidirectional links.

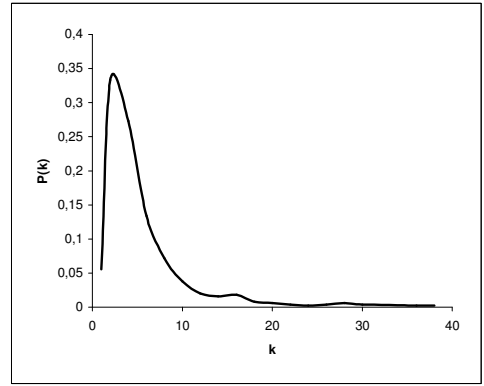

Figure 6. Degree distribution of a graph with M=5 and bidirectional links.

## 5. Evaluation of the LTM Representation

We also analyzed the structure of LTM associative network that, as expected, kept the features of a scale-free graph (Tab. 1). The system was tested enabling the retrieval of information from LTM and the analysis was repeated 30 times computing the coherence rate of the final knowledge representations.

### Table 1. LTM with 40 nodes

| M | Average path length | Average degree | Clustering coefficient |
|---|---|---|---|
| 1 | 2.56 | 5.95 | 0.32 |
| 2 | 2.49 | 6.50 | 0.34 |
| 3 | 2.27 | 8.30 | 0.45 |
| 4 | 2.25 | 9.50 | 0.43 |
| 5 | 2.23 | 9.85 | 0.43 |

The coherence rate is obtained by correlating the LTM ratings given for each item in a pair with all of the other concepts[2].

The average coherence rate (0.45) has confirmed that the conceptualization, i.e. the evolution of the associative network, was made by the system on the basis of a precise inner schema.

To evaluate the correctness of this schema we compared the LTM associative network with representations obtained from a group of human subjects. All the participants had college and university degrees and they were aged between 25 and 32.

The subjects were asked to read the same medical article examined by the system, assigning a rate of relatedness to each pair of words that were considered by the system.

---

[2] This rate was computed by the software PCKNOT 4.3, a product of Interlink Inc.

A Pathfinder analysis (R.W. Schvaneveldt, F.T. Durso, D.W. Dearholt, 1985) was performed on the relatedness matrices provided by human subjects and the LTM matrix, in order to extract the so called "latent semantics", i.e. other implicit relations between words. The obtained matrices were compared using a similarity rate determined by the correspondence of links in each pair of networks.

The first two layouts (figure 7 and 8) refer to the representation obtained by the system with and without the information retrieval from the LTM respectively.

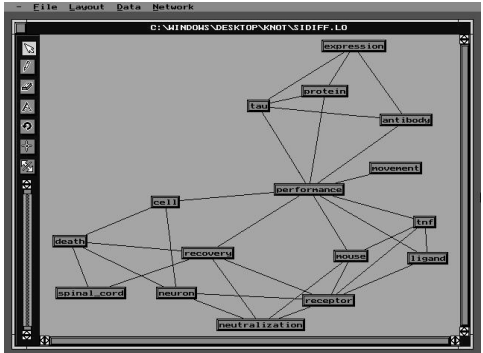

**Figure 7. LTM representation with the diffusion of the activation signal.**

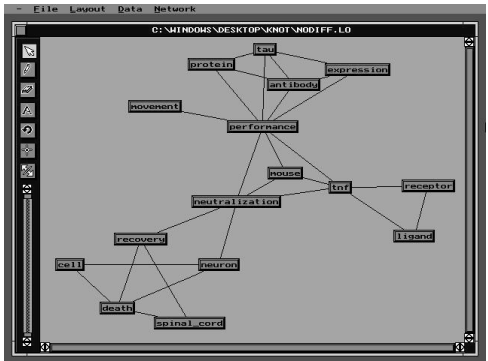

**Figure 8. LTM representation without the diffusion of the activation signal.**

The coherence rate of the network that was obtained with the diffusion of the activation signal is lower than the other one due to the fact that more links are created (31 links vs 27 links). The two networks have 23 common links and the similarity rate of the two representations is very high (0.657). But the first representation presents more significant hubs.

With the information retrieval from the LTM we have the following hubs : performance (degree 9), receptor (degree 6), recovery (degree 5). We must consider that the article deals with the functional recovery by the ligand/receptor systems. So in the first network the hubs are very significant.

Without the information retrieval we have : performance (degree 8), tnf and neutralization (degree 5) that are less significant words.

If only a scale free model or a self organizing network is used to represent the content of a text, its structure converges rapidly towards a state which has an high coherence rate.

The information retrieval from LTM during the analysis of a new text breaks this coherence and the entire system can be considered to be an information amplifier that leads to a continuous emergence of new cognitive structures. It has to be underlined that the expanding of the semiotic domain regards the user-network couple considered as a whole system.

The intrinsic emergence of the system shows itself with the appearance of new codes represented by the structure of the links of the networks after the analysis of each new document.

So the user, by the selection of the documents that must be processed by the system, can indirectly affect the cognitive process which characterizes the system knowledge representation.

Three of the 21 experimental data were discarded for their low coherence rate (below 0.2).

The others were subdivided in three groups whose characteristics are reported in table 2.

**Table 2. characteristics of the group representations**

| Group no. | Average coherence rate | Average similarity rate | Similarity rate |
|---|---|---|---|
| 1 | 0.780 | 0.271 | 0.243 |
| 2 | 0.698 | 0.218 | 0.211 |
| 3 | 0.380 | 0.162 | 0.122 |

The first two reported parameters are the average coherence rate and the average similarity rate of the representations belonging to each group. The last parameter is the similarity rate between the representation of our system and the three group representations that were obtained combining and averaging the relatedness rates provided by the members of each group.

Figures 9-11 show the resulting networks of groups 1-3.

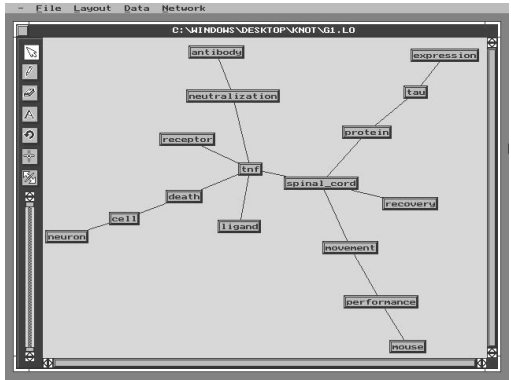

Figure 9. Group 1 representation.

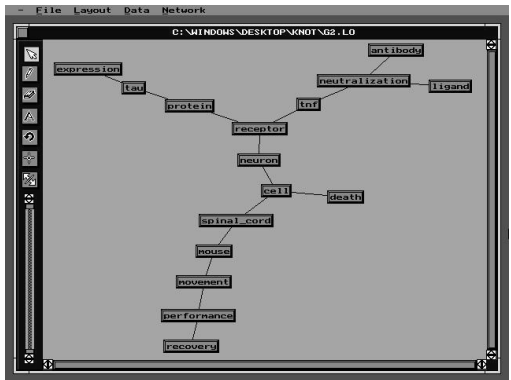

Figure 10. Group 2 representation.

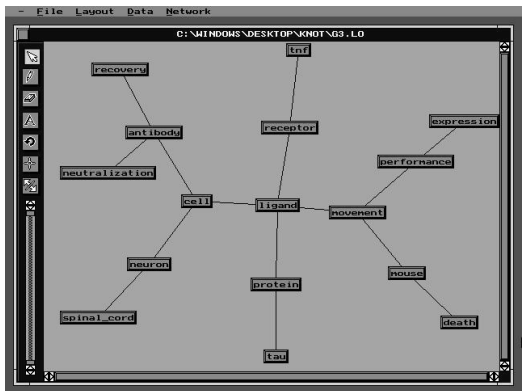

Figure 11. Group 3 representation.

It can be noticed that the first group representation presents two fundamental hubs (spinal chord, tnf) and it has the highest coherence rate.

The second group representation has a lower coherence rate and different hubs (cell, receptor, neutralization).

The third group representation is the most chaotic one with the lowest coherence rate. The hubs (ligand, movement, cell and antibody) are less representative than the others.

The representation of our system is more similar to the representations of the first two groups which have 9 and 8 common links (out of a total of 15) with the network of the system and the probability to have exactly this number of links in common by chance being very low (0.248%).

Now we are going to repeat the same test on a group of experts in order to evaluate the similarity rate between the system representation and their representation.

Furthermore the knowledge base of the system will be enriched by the analysis of other medical articles about spinal chord injuries. In this way we think that, having more a-priori knowledge about the article's topic, the system will be able to conceptualize better the article.

**6. Conclusion**

We have presented an innovative knowledge acquisition system based on the long term working memory model developed by Kintsch and Ericsson. The knowledge of the system is structured as an associative network that is dynamically updated by the integration of scale-free graphs that represent the content of the new analyzed documents. By the diffusion of an activation signal in the LTM associative network, all the information necessary to identify the context of the analyzed concepts (terms) is retrieved.

The analysis of the WM and LTM networks have confirmed that they are both examples of scale-free graphs. The computation of the coherence rate of LTM networks revealed that the system acquires knowledge on the basis of precise inner schema whose correctness was evaluated by the comparison with the other associative networks obtained from human subjects.

Certainly our system is susceptible to improvements. Maybe the presence of the external feedback of the human user could help the system to model correctly its knowledge. For example the links in the LTM could be strenghten only when the knowledge representation is used to filter or retrieve documents correctly. Furthermore the association of an age to the links of the LTM could guarantee more plasticity to its structure. This further information could be used in the computation of the fitness values as in the Dorogovtzev models (Dorogovtsev and Mendes, 2000).

The structure of the associative network can be extended using the content of an ontology based on the definitions of WordNet in order to obtain a network of propositions representable by the languages of the Semantic Web.

This ontology is created dynamically considering the context i.e. the content of the analyzed text and the nodes representing the goals and the requests of the user (query).

This is an innovative technique that could represent the first concrete ontogenetic approach to the problem of the knowledge acquisition.